\documentclass{PoS}
\usepackage{graphicx}

\newcommand{\bee}{\begin{equation}}
\newcommand{\ee}{\end{equation}}
\newcommand{\beea}{\begin{eqnarray}}
\newcommand{\eea}{\end{eqnarray}}
\def\Tr{{\rm Tr}\,}
\def\Re{{\rm Re}\,}

\title{Gauge theories with fermions in the two-index symmetric representation}

\ShortTitle{Fermions in symmetric representation}

\author{\speaker{Thomas DeGrand}\\
Department of Physics,
University of Colorado, Boulder, CO 80309, USA\\
E-mail: \email{degrand@pizero.colorado.edu}}

\author{Yigal Shamir and Benjamin Svetitsky\\
Raymond and Beverly Sackler School of Physics and Astronomy,
Tel~Aviv University, 69978
Tel~Aviv, Israel\\
Email: \email{shamir@post.tau.ac.il}, \email{bqs@julian.tau.ac.il}}

\abstract{
We summarize our recent work on gauge theories with two flavors of  fermions in
 the two-index symmetric representation:
SU(2) gauge theory with adjoint fermions, SU(3) with sextets, and SU(4) with
 ten-dimensional-representation
fermions. All three systems have beta functions smaller than their perturbative value, approaching
a fixed point near the expected two-loop zero. In all cases the mass anomalous dimension is small, under 0.5.
}

\FullConference{ The XXIX International Symposium on Lattice Field Theory - Lattice 2011\\
July 10-16, 2011\\
Squaw Valley, Lake Tahoe, California}

\begin{document}

For the last several years we have been studying
 SU($N$) gauge theories
with two flavors of two-index symmetric-representation fermions.
These theories have been proposed as candidate models for walking technicolor~\cite{Hill:2002ap}.
To supply the phenomenology of technicolor, the theories must
be confining and chirally broken, so that they possess Goldstone bosons to be eaten
by the $W$ and~$Z$. To supply the phenomenology of {\em extended\/} technicolor, that is, to
generate phenomenologically viable values for quark masses without simultaneously
generating too-large flavor-changing neutral currents, requires a mass anomalous
dimension $\gamma_m$, defined by
\bee
\mu \frac{dm(\mu)}{d\mu} = -\gamma_m(g^2) m(\mu),
\ee
that is large, of order unity. Technicolor models that produce both electroweak symmetry breaking
and fermion mass generation are called ``walking technicolor.''
The coupling is presumed to run to a large value, at which $\gamma_m$ is
large, and then to stall for many decades due to a near-zero value of the
beta function, till finally chiral symmetry breaking sets in.
Since the beta function and $\gamma_m$ are
 the important ingredients
in the phenomenology, our work has focused on measuring them. We do this using Schr\"odinger-functional
background-field
techniques.

The status of this project is as follows:
\begin{itemize}
\item We observed an infrared-attractive fixed point (IRFP) in SU(2) with two
flavors of adjoint-representation fermions,
and we have measured $\gamma_m$. This is published~\cite{DeGrand:2011qd}.
\item Last year \cite{DeGrand:2010na}
we published a result for a beta function for SU(3) with two flavors
of sextet fermions, which became small
in strong coupling. We could not
push to stronger coupling because of the presence of a phase transition that is a lattice artifact. With  new techniques  (to be described below) the beta function is consistent
with zero at the strongest coupling we reach.
\item  We are also simulating SU(4) gauge theory with two flavors
of ten-dimensional fermions. Its beta function also falls from its one-loop
perturbative value to zero at our strongest couplings.
\end{itemize}
In all cases the mass anomalous dimension $\gamma_m(g^2)$ is small, less than 0.5
 over the observed range.
The data we presented at the conference have been updated to the time this report is being written.
The SU(3) and SU(4) analyses, however, are still incomplete.

\medskip

In the Schr\"odinger
functional, the running coupling is defined by the response of the effective action to the boundary conditions.
 The scale $L$ is given by the size of the simulation volume and a scale change $s$ is achieved by
performing simulations at several values ($L$, $sL$) of the volume at fixed bare couplings.
Theories with many fermion degrees of freedom are characterized by slow running of the
effective coupling constant.
This can be seen even at one loop:
\bee
b(1/g^2)\equiv\frac{d(1/g^2)}{d\log L}  = 2\frac{\beta(g^2)}{g^4}  = 2\frac{b_1}{16\pi^2} +
\dots
\ee
where $b_1 = -\frac{11}{3}N_c + \frac{4}{3}N_f T(R)$. The coupling then runs as
\bee
\frac{1}{g^2(s)} = \frac{2b_1}{16\pi^2} \log s + \dots
\label{eq:lobf}
\ee
Now the range of scales $s$ accessible to a set of lattice simulations at any value of the bare parameters
is small, certainly $s<10$.
If $b_1$ is small, than over this range of scales the one-loop coupling will scarcely change;
the system will behave as if it is
nearly conformal.
This behavior is expected at weak coupling, near the Gaussian fixed point.
At stronger coupling, the beta function can decrease
in absolute value, in which case the theory is even more nearly conformal; or it could increase,
as in ordinary low-$N_f$ QCD. In all the cases we have studied, the beta function decreases toward zero
as we move to stronger coupling.

This slow running has both good and bad consequences for a simulation.
A good consequence is that the data become easy to analyze. For example, correlation functions
 in a near-conformal theory become pure power laws,
\bee
\Gamma(sp) = s^{d_n} \Gamma(p) \exp \int_1^s \frac{dt}{t}\gamma(g(t))
   \simeq s^{d_n} \Gamma(p) s^{\gamma(g(s))} .
\label{eq:slowg}
\ee
A negative consequence is that if we want a system to be strongly interacting at long distance,
it must also be strongly interacting at short distance. This means that one's simulation
can be strongly affected by discretization artifacts in the lattice action.

We fight lattice artifacts by using improved actions.
Nearly all our studies use clover fermions, with ``fat link'' gauge connections, specifically
nHYP links~\cite{Hasenfratz:2001hp,Hasenfratz:2007rf}.
 These fermions have excellent scaling properties when used in conventional QCD simulations.
The good scaling behavior we observe in these studies justifies their use {\em a posteriori}.

To carry out a Schr\"odinger functional study of a running coupling, we must simulate
at zero fermion mass. This amounts to simulating along
the $\kappa_c(\beta)$ line in bare parameter space.
Because Wilson-type fermions suffer an additive mass renormalization, we
determine the fermion mass through the axial Ward identity (AWI).
Systems with Wilson-type fermions and many fermion degrees of freedom
(many fundamental flavors or a few flavors of higher dimensional fermions)
have an annoying first-order strong-coupling phase transition.
Across this transition the AWI quark mass
is discontinuous, and it jumps from a positive to a negative value.
Thus there is no place where the fermions are massless in strong coupling:
The $\kappa_c$ line just comes to an end. If we are to observe an IRFP,
it must lie on the part of the $\kappa_c$ line which is not masked by the strong coupling transition.

The transition is a lattice artifact. Different lattice actions can move it around.
 For SU(2) with $N_f=2$ adjoints, replacing the thin-link Wilson action,
used by all earlier studies~\cite{Catterall:2007yx,Catterall:2008qk,Hietanen:2009az,Bursa:2009we},
with nHYP clover fermions pushed the transition back and exposed the
 IRFP\@. For SU(3) and SU(4) this change of action
was insufficient; the first order transition remained at relatively weak coupling.

To this point, all our simulations had been done with the plaquette gauge action.
After some trial and error
 we discovered that if we changed the gauge action, we could push
the transition back, and study stronger coupling.
We did this by supplementing
the original plaquette term
with an additional plaquette term, constructed with the same link as is used in the fermion action---a fat link in the higher representation,
\beea
S_G &=&
\frac{\beta}{2N} \sum \Re \Tr U_\mu(x) U_\nu(x+\hat\mu) U_\mu^\dagger(x+\hat\nu) U_\nu^\dagger(x)
\nonumber\\
& &+ \frac{\beta_f}{2d_f} \sum \Re \Tr V_\mu(x) V_\nu(x+\hat\mu) V_\mu^\dagger(x+\hat\nu) V_\nu^\dagger(x) \nonumber \\
\eea
where $U_\mu(x)$ is the thin link, $V_\mu(x)$ is the fat link, $N$ is the number of colors,
and $d_f$ is the dimensionality of the fermion representation.
In lowest order, this action is just a quadratic form in the vector potentials of the thin and fat links.
Purely empirically, we found that a positive $\beta_f$ does the job. We did most of our tests for
SU(4), where we settled on $\beta_f=1.0$.
 For SU(3) we settled on   $\beta_f=0.5$ and used it without extensive tests.

\section{SU(2)}

Now we go on to our results. We took data at  lattice sizes $L=6$, 8, 12, 16.
In these slowly running theories, a plot of the inverse gauge coupling versus $\log L$ at fixed $(\beta,\beta_f,\kappa)$
is  essentially a straight line, whose
slope  is the beta function $ b(1/g^2)$ for the inverse coupling.
We determine the slope $b(1/g^2)$ from a simple linear fit, and  check for
potential $a/L$ discretization effects by dropping the $L=6$ point from the fit.

\begin{figure}
\vspace*{20ex}
\begin{center}
\begin{picture}(200,50)(0,0)
\put(-70,0){\includegraphics[width=0.35\textwidth]{1g2_vs_l_su2.eps}}
\put(7,125){{\tt (a)}}
\put(110,-4){\includegraphics[width=0.355\textwidth]{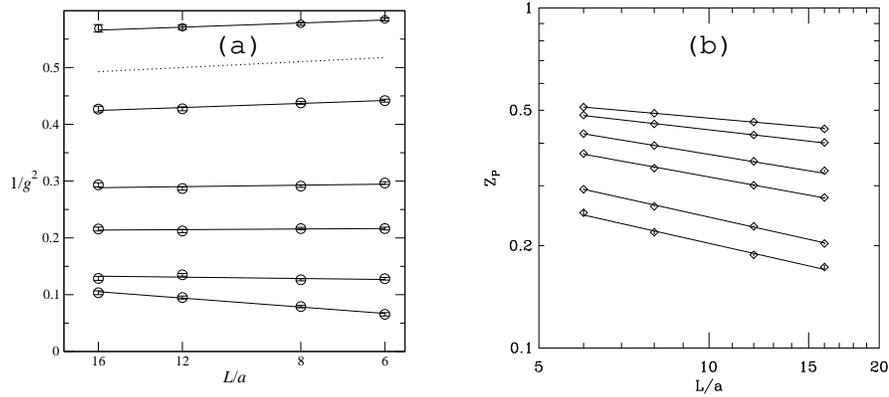}}
\put(185,125){{\tt (b)}}
\end{picture}
\end{center}
\caption{
(a) SF inverse coupling $1/g^2(L)$ vs $\log L$ in the SU(2) theory.
Lines are linear fits to data at fixed bare coupling.
The beta function is the slope of each line.  The isolated dotted line is the AF slope.
(b) Pseudoscalar renormalization constant $\log Z_p$ vs $\log L$
for the same simulation points.  $\gamma_m$ is minus the slope of each line.
\label{fig:vslsu2}}
\end{figure}

As usual
in Schr\"odinger functional calculations~\cite{Sint:1998iq,Capitani:1998mq,DellaMorte:2005kg,Bursa:2009we},
we obtain $\gamma_m$ from the renormalization factor $Z_P$ of the pseudoscalar density.
Taking advantage of the slow running as in Eq.~\ref{eq:slowg}, we fit
$\log Z_P(L) = -\gamma_m \log L +$ constant. Dropping the $L=6$ point or doing more
 complicated fits allows us to search for
lattice artifacts.
A detailed discussion  of our fitting methodology for SU(2)  may be found in
  Ref.~\cite{DeGrand:2011qd}.
We illustrate our raw data with results from the SU(2) study shown
in Fig.~\ref{fig:vslsu2}.

\begin{figure}
\begin{center}
\includegraphics[width=0.8\textwidth,clip]{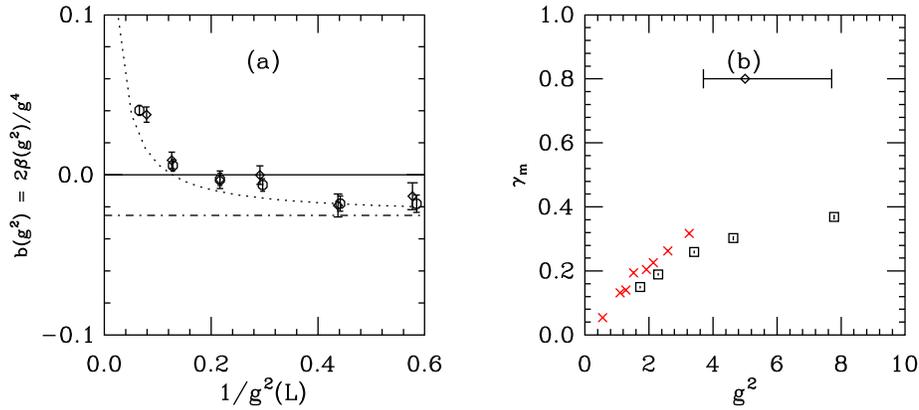}
\end{center}
\caption{ (a) Beta function for the
inverse SF coupling in the SU(2) theory.
The curves are the lowest-order and two-loop beta functions.
Squares are fits to $L\ge 6$, octagons to $L\ge 8$.
(b)
Mass anomalous dimension $\gamma_m(g^2)$ from
a linear fit to the $\log Z_p$ vs $\log L$ data, shown as squares.
The horizontal bar at the top marks our result
for $g_*^2$.
The crosses are the data of Bursa {\em et al.}~\protect{\cite{Bursa:2009we}}\@,
analyzed with the same linear fit.
\label{fig:betasu2}}
\end{figure}

 Fig.~\ref{fig:betasu2} displays our results for the beta function and $\gamma_m(g^2)$.
 Note how
 the mass anomalous dimension reaches a plateau at strong coupling.
 Even though our determination of the IRFP $g_*^2$ has large uncertainty,
 the weak dependence of $\gamma_m$ on
 $g^2$ allows for a tight determination of the anomalous dimension at the IRFP,
 $\gamma_m(g_*^2)=0.31(6)$.

\section{SU(3)}
\begin{figure}
\begin{center}
\includegraphics[width=0.8\textwidth,clip]{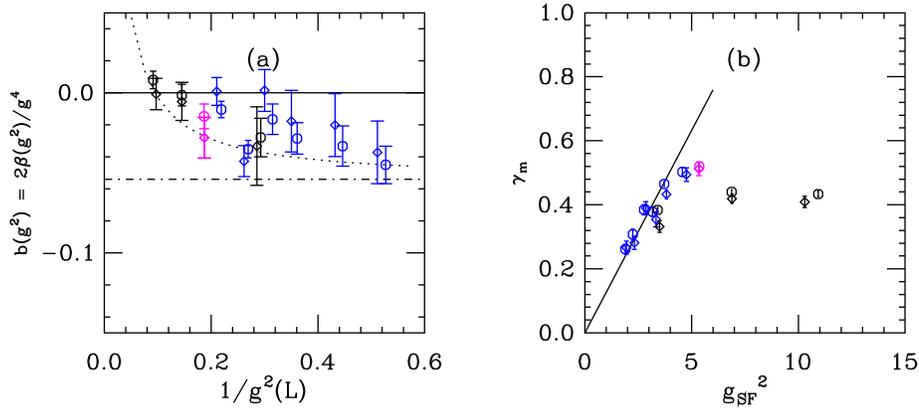}
\end{center}
\caption{ (a) Beta function for the
inverse SF coupling in the SU(3) theory. Octagons are fits to the slopes of the lines  $1/g^2(L)$ vs $\log L$.
Diamonds drop the $L=6$ points from the fit.
The curves are the lowest-order and two-loop beta functions.
Blue and pink data points are from Ref.~\protect{\cite{DeGrand:2010na}}, obtained with $\beta_f=0$; the pink points
were obtained in a metastable phase, past the strong coupling transition.
The black points are recent data, obtained with $\beta_f=0.5$.
(b)  Mass anomalous dimension vs
SF coupling, colors as in (a). The line is the one-loop result.
Where they disagree, the black points supersede the blue points,
which are influenced by the proximity of the lattice-artifact first-order
transition.
\label{fig:allsu3}}
\end{figure}

The SU(3) gauge theory coupled to two flavors of sextet quarks has been the
subject of most of our research, from a study of spectroscopy~\cite{DeGrand:2008kx} to the beta function and anomalous dimension via Schr\"odinger functional~\cite{Shamir:2008pb} and finite-size scaling~\cite{DeGrand:2009hu}.
 An early study with small volumes using a thin-link clover
action, which indicated an IRFP~\cite{Shamir:2008pb}, was superseded by a set of
simulations~\cite{DeGrand:2010na} on larger volumes using fat links.
The latter did show
the beta function running to a small value at strong coupling, but no IRFP.
The strong coupling transition prevented us from pushing further into strong coupling. With the new two-term
gauge action, we can reach a coupling close to that of the
two-loop Banks--Zaks~\cite{Banks:1981nn} fixed point. At this point, we find that the beta function is consistent with zero and probably crosses zero.
As in the SU(2) theory, the mass anomalous dimension follows the perturbative value out of weak coupling,
until it breaks away and becomes independent of $g^2$, taking a value under 0.5.
See
Fig.~\ref{fig:allsu3}.

\section{SU(4)}
\begin{figure}
\begin{center}
\includegraphics[width=0.8\textwidth,clip]{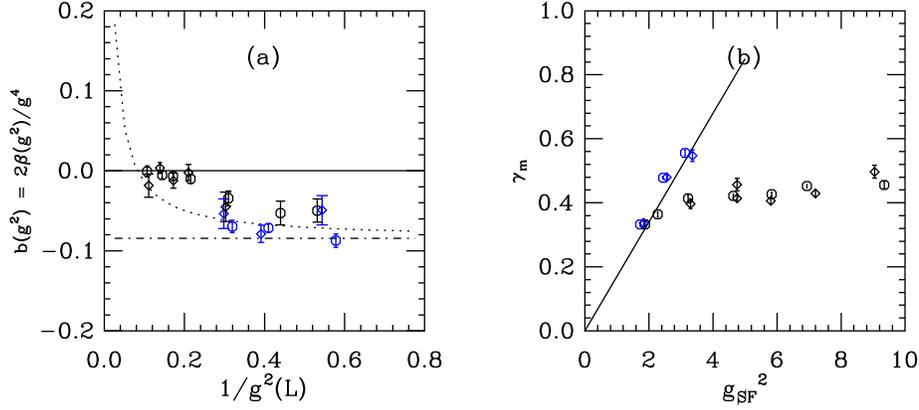}
\end{center}
\caption{ (a) Beta function for
inverse SF coupling for $\beta_f=0$ and~1.0 (blue and black points,
respectively), in the SU(4) theory. Plot symbols are as for SU(3).
The curves are the lowest-order and two-loop beta functions.
(b) Mass anomalous dimension vs
SF coupling  for $\beta_f=0$ and~1.0.
The line is the one-loop result.  Once again,
the $\beta_f=1.0$ results supersede the $\beta_f=0$ results where they disagree.
\label{fig:allsu4}}
\end{figure}

This year we studied the third of our related theories, SU(4) gauge theory coupled
to two flavors of ten-dimensional fermions. Even with nHYP links in the fermion action, the
strong-coupling transition was encountered at quite a weak coupling. The
fat-link gauge action, however, allows us to push farther into strong coupling. As in the case of SU(3),
 our strongest-coupling points
are consistent with a zero beta function.
(We believe that with our present actions we will be unable
to push to stronger coupling because of low acceptance.)
Again, the zero is at slightly weaker coupling than the
Banks--Zaks point. The mass anomalous dimension again falls off the perturbative
curve to take a nearly $g^2$-independent value, under 0.5, in strong coupling.
Fig.~\ref{fig:allsu4} shows these results.

\section{Conclusions}

The analysis of the SU(3) and SU(4) systems is still in progress. Nevertheless,
the outline of our conclusion is clear: Over the range where we can perform simulations
the mass anomalous dimension never exceeds 0.5 in any of the three models.
This strongly disfavors them as candidate theories for walking
(extended) technicolor.
As far as we can tell, all three models exhibit an IRFP at a value of Schr\"odinger functional
coupling slightly weaker than expected by two-loop perturbation theory.
To conclude on a positive note, these theories give theorists a set of ``tame'' lattice-regulated
gauge theories with infrared-attractive fixed points suitable for additional
studies by numerical simulation.

\acknowledgments

B.~S. and Y.~S. thank the University of Colorado for hospitality.
This work was supported in part by the Israel Science Foundation
under grant no.~423/09 and by the U.~S. Department of Energy.
Computations were carried out at the University of Texas and at the National Institute for Computational Sciences (NICS) at the University of Tennessee,
 through TeraGrid/XSEDE grants no.~TG-PHY080042 and no.~TG-PHY090023 funded by the National Science Foundation.
Additional computations were done on clusters at the University of Colorado and Tel Aviv University,
as well as on facilities of the USQCD Collaboration at Fermilab,
which are funded by the Office of Science of the U.~S. Department of Energy.
Our computer code is based on the publicly available package of the
 MILC collaboration~\cite{MILC}.

\end{document}